\documentclass[aps,prl,showpacs,twocolumn]{revtex4}

\usepackage{graphicx}
\usepackage[ansinew]{inputenc}
\usepackage{amsmath,amsfonts,amssymb}

\begin{document}

\bibliographystyle{prsty}

\title{Quantized Conductance of a Single Magnetic Atom}
\author{N. N\'{e}el}
\author{J. Kr\"{o}ger}\email{kroeger@physik.uni-kiel.de}
\author{R. Berndt}
\affiliation{Institut für Experimentelle und Angewandte Physik,
Christian-Albrechts-Universität zu Kiel, D-24098 Kiel, Germany}
\date{\today}

\begin{abstract}
A single Co atom adsorbed on Cu(111) or on ferromagnetic Co islands is contacted
with non-magnetic W or ferromagnetic Ni tips in a  scanning tunneling
microscope. When the Co atom bridges two non-magnetic electrodes conductances
of $2\,\text{e}^2/\text{h}$ are found. With two ferromagnetic electrodes a
conductance of $\text{e}^2/\text{h}$ is observed which may indicate fully
spin-polarized transport.
\end{abstract}

\pacs{68.37.Ef,72.25.Ba,73.23.Ad,73.63.-b,73.63.Rt}

\maketitle

The conductance of nanometer-sized contacts may be decomposed into contributions
of transport eigenchannels according to $G=\text{G}_0\sum_{i=1}^{n}\tau_i$,
where $\text{G}_0=2\text{e}^2/\text{h}$ is the quantum of conductance (-e:
electron charge, h: Planck's constant), and $\tau_i$ is the transmission
probability of the $i$th channel \cite{rla_57,mbu_88}. The factor $2$ in the
quantum of conductance is due to spin degeneracy. In contacts involving
magnetic electrodes the spin degeneracy of transport channels may be lifted.
Each spin-polarized channel then may contribute up to $\text{G}_0/2$ to the
total conductance.

A quantized conductance of $\text{G}_0/2$ is expected when a fully spin-polarized
current is transmitted with a probability of $1$ through a spin-polarized
transport channel. These conditions appear difficult to fulfill. Nevertheless,
experimental observations of conductance quantization in units of $\text{G}_0/2$
have repeatedly been reported \cite{jco_97,ton_99,fel_02,msh_02,vro_03,dgi_03}.
These conductances were observed with \cite{ton_99,msh_02} or without
\cite{jco_97,fel_02,vro_03,dgi_03} external magnetic fields, for ferromagnetic
\cite{jco_97,ton_99,fel_02,msh_02,vro_03} and non-magnetic electrodes
\cite{vro_03,dgi_03}. On the other hand, the absence of non-integer conductance
quantization has also been inferred from experimental results \cite{csi_96,cun_04}.
Untiedt {\it et al.}\ showed that contaminants like $\text{H}_2$ or CO modify
the conductance and could, in the case of CO adsorption on Pt electrodes, give
rise to a conductance of $\text{G}_0/2$ \cite{cun_04}. A considerable variety
of model calculations \cite{aba_04,dja_05,asm_06,aro_07,mha_08} have been
performed and support the notion that conductance quantization in units of
$\text{G}_0/2$ is not expected from the investigated ferromagnetic contacts.
It should be noted, however, that the modeling performed so far did not include
geometrical relaxations of the contacts although the importance of the detailed
atomic arrangement has been emphasized \cite{aba_04,mha_08,jcu_98,jve_04}.

The contradictory conclusions reached from the various experiments may be
related to a lack of characterisation of the atomic details of the junction.
This problem may be reduced by using a cryogenic scanning tunneling microscope
to probe the conductance of clean single-atom contacts in ultra-high vacuum.
Here we apply this approach to investigate prototypical junctions. A single
magnetic atom on a spin-polarized island or a non-magnetic substrate is
contacted with non-magnetic and ferromagnetic tips. We find that the conductance
of a single Co atom is $\text{G}_0$ when two non-magnetic electrodes are used.
With ferromagnetic electrodes the conductance is $\text{G}_0/2$. Conductances
of $\approx 0.9\,\text{G}_0$ are observed for a combination of a non-magnetic
and a ferromagnetic electrode. In contrast to previous experiments, the contact
geometry and chemistry are characterised by imaging of the contact area prior
to and after conductance measurements. We hint that the observed conductance
may be related to the detailed geometry and bonding at the contact.

The experiments were performed using a home-built scanning tunneling microscope
operated at $7\,\text{K}$ and in ultra-high vacuum with a base pressure of
$10^{-9}\,\text{Pa}$. The Cu(111) surface as well as chemically etched W and
Ni tips were cleaned by argon ion bombardment and annealing. Tungsten and Ni
tips were fabricated from $0.25\,\text{mm}$ thick wire of $99.99\,\%$ purity.
Cobalt deposition onto Cu(111) was performed at room temperature using an
electron beam evaporator and a Co evaporant of $99.99\,\%$ purity. Single Co
atoms were deposited onto the cold sample surface through openings in the
cryostat shields. The tip magnetization direction is dictated by the shape
anisotropy since magnetocrystalline anisotropy is small in Ni. Therefore, the
tip is magnetized along its axis leading to a magnetization perpendicular to
the substrate surface \cite{mra_06}. Cobalt islands on Cu(111) are well studied
\cite{jfi_93,mop_97,opi_04} and identified as single-domain ferromagnetic
exhibiting a perpendicular magnetization with strong coercivity and remanence
\cite{opi_04}.

\begin{figure}[t]
  \includegraphics[width=85mm,clip=]{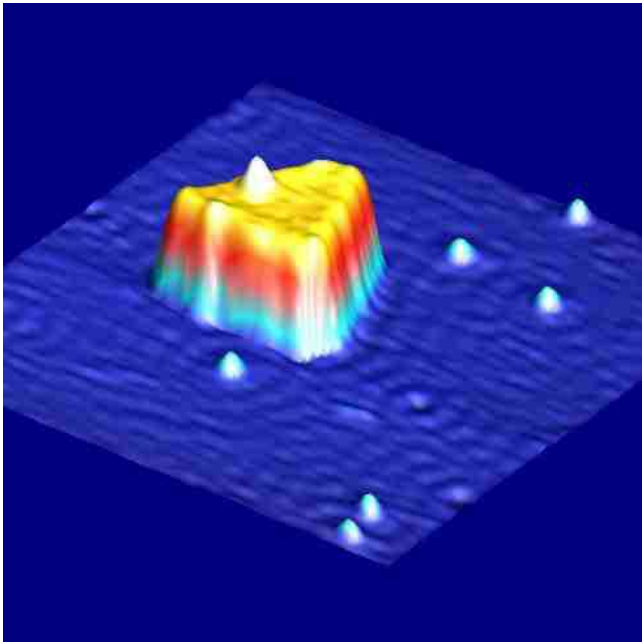}
  \caption[fig1]{(Color online) Pseudo-three-dimensional representation of
  a constant-current STM image of a triangular Co island adsorbed on Cu(111).
  Single Co atoms that are adsorbed on the substrate surface and on top of
  the island appear as protrusions (sample voltage: $V=100\,\text{mV}$,
  current: $I=100\,\text{pA}$, size: $24.5\,\text{nm}\times 24.5\,\text{nm}$).}
  \label{fig1}
\end{figure}

Figure \ref{fig1} shows a pseudo-three-dimensional scanning tunneling microscopy
(STM) image of a Co island and single Co atoms adsorbed on Cu(111) at $7\,{\rm K}$.
The Co island exhibits a triangular shape and a thickness of two Co layers as
expected. On top of the island a single Co atom was adsorbed. These surface
structures together with a non-magnetic W or a ferromagnetic Ni tip provide
four contact configurations. The W or Ni tip may contact a Co atom adsorbed
on the non-magnetic substrate surface or on a ferromagnetic Co island.

Cleanliness of the tip as well as of the Cu(111) surface and the surface of
adsorbed Co islands was monitored by spectroscopy of the differential conductance
($\text{d}I/\text{d}V$). The Shockley-type surface state of Cu(111) was observed
as a sharp step-like onset of $\text{d}I/\text{d}V$, while Co islands exhibited
occupied as well as unoccupied $d$ states as pronounced peaks in spectra of
$\text{d}I/\text{d}V$ \cite{opi_04}. Therefore, the presence of contaminants,
in particular of hydrogen, adsorbed on the substrate or the islands can be
safely ruled out. Moreover, hydrogen adsorbed on Co islands has recently
been shown to desorb by using tunneling currents exceeding $20\,\text{nA}$
\cite{msi_08}. In the contact experiments reported here, currents of the order
of $10\,\mu\text{A}$ are passed through the islands.

\begin{figure}[t]
  \includegraphics[width=85mm,clip=]{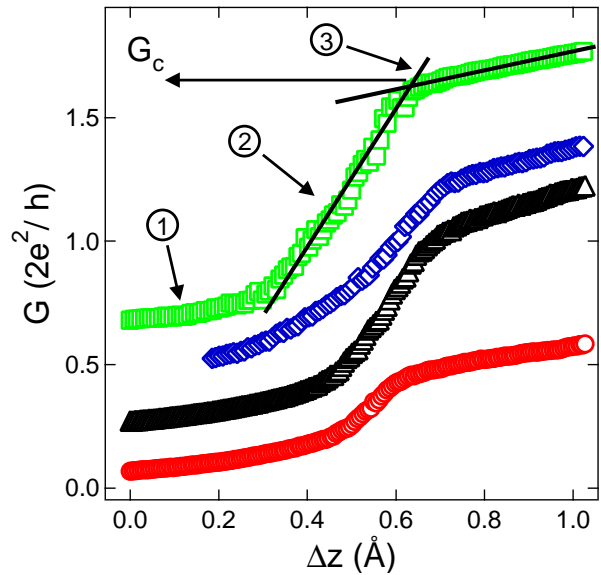}
  \caption[fig2]{(Color online) Conductance $G$ vs.\ tip displacement $\Delta z$
  recorded from a single Co atom with different electrode combinations. W or
  Ni tips are used to contact a single Co atom on a Cu substrate or on a Co
  island. Squares: W -- Co -- Cu, lozenges: W -- Co -- Co, triangles:
  Ni -- Co -- Cu, circles: Ni -- Co -- Co.  Tunneling ($1$), transition ($2$)
  and contact ($3$) regions of the conductance curves are indicated. Linear
  fits to the data in the transition and contact regions are used to define
  a contact conductance $G_{\text{c}}$. $\Delta z=0$ corresponds to
  $V=100\,\text{mV}$ and $I=500\,\text{nA}$ prior to opening the feedback
  loop of the microscope. The conductance curves for the W -- Co -- Cu,
  W -- Co -- Co, and Ni -- Co -- Cu contacts were vertically offset by
  $0.6\,\text{G}_0$, $0.4\,\text{G}_0$, and $0.2\,\text{G}_0$, respectively.}
  \label{fig2}
\end{figure}

Figure \ref{fig2} displays the conductance of a single Co atom between different
combinations of electrodes, as a function of the displacement $\Delta z$ of
the microscope tip. For small tip displacements the conductance varies exponentially
with the displacement as expected for the tunneling regime (denoted $1$ in
Fig.\,\ref{fig2}). In a transition region ($2$) the conductance rapidly increases.
Finally, a smaller variation of  the conductance occurs in the contact region
($3$). To define a contact conductance, $G_{\text{c}}$, we approximate the
conductance data in the transition and contact regions by straight lines.
Their point of intersection defines the contact conductance. This definition
has previously been used for contacts to noble metal atoms and molecules and
was found to reproduce their expected contact conductances \cite{jkr_08}.
Conductance values obtained from Co atoms according to this procedure are
summarized in Table \ref{tab1}.

While for the W -- Co -- Cu junction the contact conductance is $\approx\text{G}_0$,
which is in agreement with single-Co conductance measured on Cu(100) \cite{nne07a},
we observe lower contact conductances with ferromagnetic electrodes. The most
striking result is obtained when a Co atom adsorbed to a Co island is contacted
by a ferromagnetic tip. In this case the contact conductance is $\approx\text{G}_0/2$.
Combinations of non-magnetic and ferromagnetic electrodes lead to Co atom
conductances of $0.88\,\text{G}_0$ (W -- Co -- Co) and $0.85\,\text{G}_0$
(Ni -- Co -- Cu).

\begin{table}
  \caption{Contact conductances $G_{\text{c}}$ of a single Co atom for
  different electrode combinations. Uncertainty margins reflect the standard
  deviations of the contact conductances observed in repeated experiments}
  \begin{ruledtabular}
    \begin{tabular}{cc}
      Tip--Atom--Surface Materials & $G_{\text{c}}\,(\text{G}_0)$ \\
      \colrule
      W -- Co -- Cu                & $1.03\pm 0.02$               \\
      W -- Co -- Co                & $0.88\pm 0.02$               \\
      Ni -- Co -- Cu               & $0.85\pm 0.02$               \\
      Ni -- Co -- Co               & $0.48\pm 0.02$               \\
    \end{tabular}
  \end{ruledtabular}
  \label{tab1}
\end{table}

We note that conductance curves acquired for voltages $|V|\leq 0.1\,\text{V}$
exhibited the same characteristics as presented in Fig.\,\ref{fig2}. Voltages
$|V|>0.1\,\text{V}$ led to an enhanced mobility of Co atoms adsorbed on Co
islands. We experienced the Co atom to change its adsorption site - either to
the tip or to an adjacent site on the island - during tip excursion toward
the adsorbed atom.

These results clearly show that the conductance of a ferromagnetic
Ni -- Co -- Co contact is close to $\text{G}_0/2$. At present, it is not clear
whether this value is due to transport through a combination of partially open
channels or, most excitingly, a single fully spin-polarized channel. In any
event, the experimental result appears to contradict current modeling results
on ballistic transport through magnetic constrictions which do not indicate
conductance quantization in units of $\text{G}_0/2$.

We suggest that relaxations of the atomic positions in the contact region may
be at the origin of this discrepancy. Recent theoretical work by H\"{a}fner
{\it et al.}\,\cite{mha_08} lends support to this interpretation. The conductance
of atomic-size Co and Ni contacts and the spin polarization of the current
were reported to be very sensitive to the contact geometry. In particular, for
electrode separations in the tunneling range, H\"{a}fner {\it et al.}\ calculated
a conductance of essentially $\text{G}_0/2$ reflecting a spin polarization of
the current of nearly $100\,\%$. For smaller electrode separations at contact,
however, the spin polarization was predicted to drop sharply to zero over a
range of $\approx 0.5\,\text{\AA}$. Relaxations of atomic positions owing to
adhesive forces, which have been found for other single-atom \cite{lli_05}
and single-molecule \cite{nne07b} contacts, were not included in these
calculations. They may shift the range of distances where spin polarization
is lost. This scenario, which remains to be analyzed by detailed calculations,
would imply that the observed $\text{G}_0/2$ conductance is due to a fully
spin-polarized channel.

In conclusion, we observed a conductance of $\text{G}_0/2$ from single Co
atoms between ferromagnetic electrodes. The contacted atom as well as the
status of the electrodes were characterized by imaging and spectroscopy of
the atom and the hosting surface. The observed conductance reduction from
$\approx 1\,\text{G}_0$ to $\approx\text{G}_0/2$ appears to contradict available
modeling results for transport through magnetic atoms. It calls for calculations
which take into account the detailed structure and bonding of the junction.

We thank M.\ Brandbyge (Technical University of Denmark) for preliminary
calculations and discussions. Financial support by the Deutsche
Forschungsgemeinschaft through SFB 668 is acknowledged.


\begin{thebibliography}{nnnyys}

  \bibitem{rla_57} R. Landauer,
  IBM J.\ Res.\ Dev.\ {\bf 1}, 223 (1957).

  \bibitem{mbu_88} M. B\"{u}ttiker,
  IBM J.\ Res.\ Dev.\ {\bf 32}, 63 (1988).

  \bibitem{jco_97} J. M. Costa-Kr\"{a}mer,
  \prb {\bf 55}, R4875 (1997).

  \bibitem{ton_99} T. Ono, Y. Ooka, H. Miyajima, and Y. Otani,
  Appl.\ Phys.\ Lett.\ {\bf 75}, 1622 (1999).

  \bibitem{fel_02} F. Elhoussine, S. M\'{a}t\'{e}fi-Tempfli, A. Encinas,
  and L. Piraux,
  Appl.\ Phys.\ Lett.\ {\bf 81}, 1681 (2002).

  \bibitem{msh_02} M. Shimizu, E. Saitoh, H. Miyajima, and Y. Otani,
  J.\ Magn.\ Magn.\ Mater.\ {\bf 239}, 243 (2002).

  \bibitem{vro_03} V. Rodrigues, J. Bettini, P. C. Silva, and D. Ugarte,
  \prl {\bf 91}, 096801 (2003).

  \bibitem{dgi_03} D. M. Gillingham, I. Linington, C. M\"{u}ller,
  and J. A. C. Bland,
  J.\ Appl.\ Phys.\ {\bf 93}, 7388 (2003).

  \bibitem{csi_96} C. Sirvent, J. G. Rodrigo, S. Vieira, L. Jurczyszyn,
  N. Mingo, and F. Flores,
  \prb {\bf 53}, 16086 (1996).

  \bibitem{cun_04} C. Untiedt, D. M. T. Dekker, D. Djukic, and J. M. van Ruitenbeek,
  \prb {\bf 69}, 081401(R) (2004).

  \bibitem{aba_04} A. Bagrets, N. Papanikolaou, and I. Mertig,
  \prb {\bf 70}, 064410 (2004).

  \bibitem{dja_05} D. Jacob, J. Fern\'{a}ndez-Rossier, and J. J. Palacios,
  \prb {\bf 71}, 220403(R) (2005).

  \bibitem{asm_06} A. Smogunov, A. Dal Corso, and E. Tosatti,
  \prb {\bf 73}, 075418 (2006).

  \bibitem{aro_07} A. R. Rocha, T. Archer, and S. Sanvito,
  \prb {\bf 76}, 054435 (2007).

  \bibitem{mha_08} M. H\"{a}fner, J. K. Viljas, D. Frustaglia, F. Pauly,
  M. Dreher, P. Nielaba, and J. C. Cuevas,
  \prb {\bf 77}, 104409 (2008).

  \bibitem{jcu_98} J. C. Cuevas, A. Levy Yeyati, A. Mart\'{\i}n-Rodero,
  G. Rubio Bollinger, C. Untiedt, and N. Agra\"{\i}t,
  \prl {\bf 81}, 2990 (1998).

  \bibitem{jve_04} J. Velev and W. H. Butler,
  \prb {\bf 69}, 094425 (2004).

  \bibitem{mra_06} M. V. Rastei and J. P. Bucher,
  J.\ Phys.: Condens.\ Matter {\bf 18}, L619 (2006).

   \bibitem{jfi_93} J. de la Figuera, J. E. Prieto, C. Ocal, and R. Miranda,
  \prb {\bf 47}, 13043 (1993).

  \bibitem{mop_97} M.\ {\O}.\ Pedersen, I. A. B\"{o}ricke, E. L{\ae}gsgaard,
  I. Stensgaard, A. Ruban, J. K. N{\o}rskov, and F. Besenbacher,
  Surf.\ Sci.\ {\bf 387}, 86 (1997).

  \bibitem{opi_04} O. Pietzsch, A. Kubetzka, M. Bode, and R. Wiesendanger,
  \prl {\bf 92}, 057202 (2004).

  \bibitem{msi_08} M. Sicot, O. Kurnosikov, O. A. O. Adam, H. J. M. Swagten,
  and B. Koopmans,
  \prb \textbf{77}, 035417 (2008).

  \bibitem{jkr_08} J. Kr\"{o}ger, N. N\'{e}el, and L. Limot,
  J.\ Phys.: Condens.\ Matter {\bf 20}, 223001 (2008).

  \bibitem{nne07a} N. N\'{e}el, J. Kr\"{o}ger, L. Limot, K. Palotas,
  W. A. Hofer, and R. Berndt,
  \prl {\bf 98}, 016801 (2007).

  \bibitem{lli_05} L. Limot, J. Kr\"{o}ger, R. Berndt, A. Garcia-Lekue,
  and W. A. Hofer,
  \prl {\bf 94}, 126102 (2005).

  \bibitem{nne07b} N. N\'{e}el, J. Kr\"{o}ger, L. Limot, T. Frederiksen,
  M. Brandbyge, and R. Berndt,
  \prl {\bf 98}, 065502 (2007).

\end{thebibliography}
\end{document}